# Deep Reinforcement Learning Based Volt-VAR Optimization in Smart Distribution Systems

Ying Zhang, *Student Member, IEEE*, Xinan Wang, *Student Member, IEEE*, Jianhui Wang, *Senior Member, IEEE*
Yingchen Zhang, *Senior Member, IEEE*

*Abstract*— This paper develops a model-free volt-VAR optimization (VVO) algorithm via multi-agent deep reinforcement learning (MADRL) in unbalanced distribution systems. This method is novel since we cast the VVO problem in unbalanced distribution networks to an intelligent deep Q-network (DQN) framework, which avoids solving a specific optimization model directly when facing time-varying operating conditions of the systems. We consider statuses/ratios of switchable capacitors, voltage regulators, and smart inverters installed at distributed generators as the action variables of the DQN agents. A delicately designed reward function guides these agents to interact with the distribution system, in the direction of reinforcing voltage regulation and power loss reduction simultaneously. The forward-backward sweep method for radial three-phase distribution systems provides accurate power flow results within a few iterations to the DQN environment. Finally, the proposed multi-objective MADRL method realizes the dual goals for VVO. We test this algorithm on the unbalanced IEEE 13-bus and 123-bus systems. Numerical simulations validate the excellent performance of this method in voltage regulation and power loss reduction.

## I. Introduction

ELECTRIC power systems currently adopt volt-VAR optimization (VVO) to improve operational efficiency and reduce power losses in distribution systems [1]. About 10% of the energy losses occur during transmission and distribution to customers, while 40% of the total losses occur at the distribution side, according to the U.S. Energy Information Administration [2]. Research shows that effective VVO control on various regulating devices, such as automatic voltage regulators (AVRs) and switchable capacitors, can realize voltage regulation as well as loss reduction. As a typical tool in the distribution management system (DMS), the primary goal of VVO is to keep voltages at all buses within a normal operation range, *e.g.*, 0.95~1.05 p.u., according to ANSI C84.1 standard. This topic is further motivated by the penetration of distributed generation (DG), since bidirectional power flow in active distribution systems raises the risk of voltage violation [3]. The DG units equipped with smart inverters have the flexible capability of absorbing or providing reactive power.

Thus, a VVO tool with an effective control strategy on these smart inverters can enhance the operational performance of distribution systems with DG penetration [4].

Traditionally, VVO is modeled as a mixed-integer nonlinear programming (MINLP) problem established on optimal power flow (OPF) [5]. Due to the existence of integer variables and nonlinear voltage-dependent load models in systems, the VVO formulation is *nonconvex* and *NP-hard*. Hence, earlier algorithms perform VVO using heuristics, such as [6] and [7], which do not guarantee optimality. More research converts this problem to various optimization problems, namely, mixed-integer quadratically constrained programming (MIQCQP) and mixed-integer quadratic programming (MIQP), etc. [8], [9]. For instance, [8] integrates the branch-and-bound approach to the trust-region sequential quadratic programming to iteratively solve the VVO problem. However, the iteration process is time-consuming. This low computational efficiency originates from two reasons: 1) the comprehensive modeling of various control devices largely increases the complexity of these optimization models, and 2) the combination of action variables from multiple control devices results in a huge searching space [10]. On the other hand, these studies run in a centralized manner and adopt linear or nonlinear power flow formulation for single-phase distribution systems to simplify the modeling complexity, such as [5]–[9] and [11]–[14]. However, the three-phase unbalanced operation of distribution systems is more consistent with practice.

To reduce the computational burden of these centralized algorithms, the decentralized or hierarchical methods, such as [10], [15], and [16], are used to solve VVO models in unbalanced distribution systems. For instance, dividing the feeder into several regulating zones, the authors of [10] formulated a linearized power system model to solve a zone-based optimization problem in each stage via MIQP and then performed a multi-stage coordinated operation to achieve the overall voltage regulation. Unfortunately, the iterations recorded in [10] reach up to thousands and take hundreds of seconds due to this multi-stage operation. Also, these approximation techniques may cause accuracy losses in power flow calculation and lead to suboptimal control strategies. Recently, [16] develops a bi-level VVO formulation, and the lower level models a MILP problem using a nearly linear power flow, while the upper level solves a quadratically constrained nonlinear programming (QCNP) problem based on nonlinear power flow approximation. However, [16], the same as other model-based methods mentioned above, highly depends on

Y. Zhang, X. Wang, and J. Wang are with the Department of Electrical and Computer Engineering, Southern Methodist University, Dallas, TX 75205, USA (e-mail: yzhang1@smu.edu; xinanw@smu.edu; jianhui@smu.edu).
Y. Zhang is with National Renewable Energy Laboratory, Golden, CO

442

specific optimization models and has limited capability in rapidly adapting to time-varying loads in distribution systems.

To address the limitations of these model-based approaches, recent effort applies reinforcement learning (RL) to power system operation, such as voltage control [17]–[19]. Furthermore, deep RL combining deep learning with RL is regarded as valuable alternatives to model-based methods, due to its strong exploration capability of neural networks (NNs) towards nonlinear high-dimensional searching spaces. For example, the deep RL-based methods proposed in [20] adaptively provide the voltage setpoints for generators in transmission systems. However, the existing control methods via RL, such as [19] and [20], only focus on adjusting voltage profiles but ignore the potential of VVO in power loss reduction. Also, these RL-based control methods adopt single agents and have a slower learning speed when applied to larger-scale systems with a huge searching space of variables [21]. On the other hand, they do not consider voltage-dependent loads and smart inverters installed at DG units, both of which are widely used in practical power systems [5], [16]. Also, coordinated VVO control on various regulating devices has not been investigated in three-phase unbalanced distribution systems. Hence, deep RL applications in VVO require additional effort in improving the flexibility and complexity of distribution system operation and control.

Targeting at auto-adaptive voltage control under time-varying operating conditions, we propose a data-driven and model-free VVO approach via multi-agent deep RL (MADRL) in unbalanced distribution systems. The proposed method is novel since we cast the multi-objective VVO problem for distribution systems to an intelligent deep Q-network (DQN) framework. In this framework, we consider the statuses/ratios of capacitors, AVRs, and smart inverters as action variables. These actions are determined via the agents that are trained by interacting with their environment, *i.e.*, the distribution system. The backward-forward sweep method [22] for unbalanced distribution systems provides accurate power flow results with few iterations to the environment. Moreover, by customizing a reward function that effectively guides the DQN training process, this method realizes dual goals on power loss reduction and voltage regulation simultaneously. The main contributions of this paper are threefold:

▪ We integrate multiple types of regulating devices and load models into the forward-backward sweep method for power flow calculation, which is highly efficient in unbalanced distribution systems, compared with dc and linearized ac power flow methods.

▪ Unlike the single-objective RL methods in [19] and [20] that only focus on voltage regulation and ignore the roles of the ZIP load models and smart inverters in optimizing system operation, the proposed MADRL method achieves the multi-objective VVO within milliseconds via these devices and thus can be implemented online.

▪ To further improve the computational efficiency in larger-scale three-phase systems, the proposed method assigns the global control variables to multiple DQN agents with observation sharing to handle the scalability issue effectively.

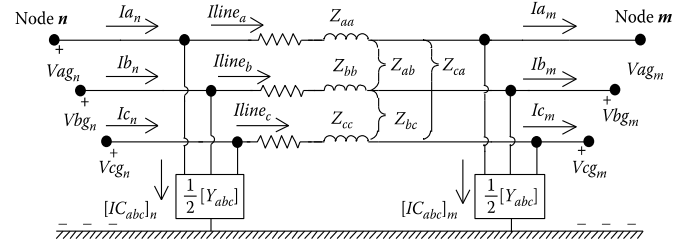

Fig.1. Three-phase line model in distribution feeders [22]

## II. System Model in Unbalanced Distribution Networks

This section introduces voltage-dependent load models and the forward-backward sweep method for power flow calculation in unbalanced distribution systems.

Unlike transmission systems, distribution networks have radial topologies with lines of a high r/x ratio, which may make the traditional Newton-Raphson power flow method fail to converge [22]. The efficient forward-backward sweep method provides exactly accurate power flow results without any approximation, even in relatively large-scale unbalanced distribution systems. Hence, this method is widely used in power flow calculation for radial distribution systems. Fig.1 shows a schematic diagram of the three-phase distribution line model between buses $n$ and $m$, where $\boldsymbol{Z}_{abc} \in \mathbb{C}^{3\times 3}$ denotes the three-phase line impedance matrix at branch $n$-$m$, and $\boldsymbol{Y}_{abc} \in \mathbb{C}^{3\times 3}$ is the shunt capacitance matrix.

We adopt the ZIP model in distribution systems, which is a voltage-dependent load model widely used in related research such as [12], [14], and [16]. The ZIP load models for active and reactive powers at bus $i$ are depicted as

$$P_i^\varphi = P_{i,0}^\varphi \left[ k_{p,1} \left(\frac{|U_i^\varphi|}{U_0}\right)^2 + k_{p,2}\left(\frac{|U_i^\varphi|}{U_0}\right) + k_{p,3} \right] \quad (1)$$

$$Q_i^\varphi = Q_{i,0}^\varphi \left[ k_{q,1} \left(\frac{|U_i^\varphi|}{U_0}\right)^2 + k_{q,2}\left(\frac{|U_i^\varphi|}{U_0}\right) + k_{q,3} \right] \quad (2)$$

where $P_i^\varphi$ and $Q_i^\varphi$ denote the $\varphi$-phase active and reactive powers at bus $i$, respectively, and $\varphi = \{a,b,c\}$; $k_{p,1} + k_{p,2} + k_{p,3} = 1$, and $k_{q,1} + k_{q,2} + k_{q,3} = 1$; $P_{i,0}^\varphi$ and $Q_{i,0}^\varphi$ denote the $\varphi$-phase active and reactive powers at the nominal voltage $U_0$; $|U_i^\varphi|$ represents the $\varphi$-phase voltage magnitude at bus $i$.

We briefly introduce the procedure of the forward-backward sweep method for power flow calculation in unbalanced distribution systems as follows:

1) *Current Injection Calculation:* Initialize three-phase voltages at all buses as the values of nominal voltages. In each iteration, the three-phase current injections at bus $k$ are calculated by

$$\begin{bmatrix} I_{k,in}^a \\ I_{k,in}^b \\ I_{k,in}^c \end{bmatrix} = \begin{bmatrix} (S_k^a/U_k^a)^* \\ (S_k^b/U_k^b)^* \\ (S_k^c/U_k^c)^* \end{bmatrix} \quad (3)$$

where $S_k^\varphi = P_k^\varphi + jQ_k^\varphi$ denotes the $\varphi$–phase complex power of load consumption or DG production at bus $k$ and is considered as the ZIP model in (1) and (2) ; $U_k^\varphi$ denotes the $\varphi$- phase voltage at bus $k$ in the current iteration, and $k = 1, 2, ..., N$; $[\cdot]^*$ represents the complex conjugate.

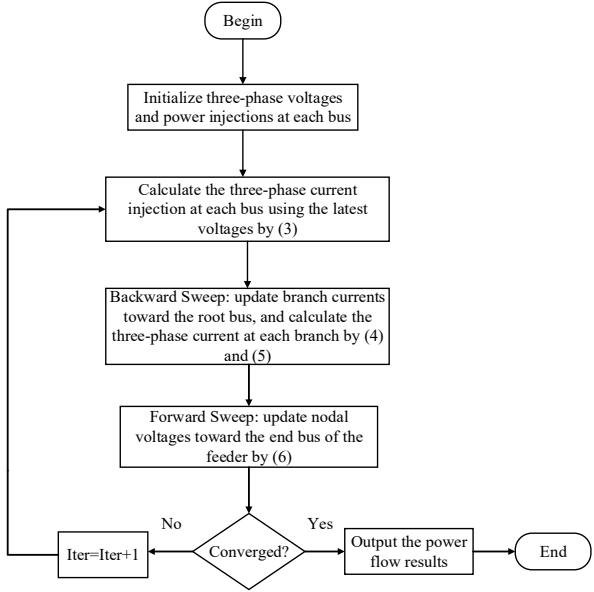

Fig.2. Flowchart of the forward-backward sweep method

2) *Backward Sweep:* Starting from the end bus of the feeder, we calculate the receiving-end current $\boldsymbol{I}_m$ at branch $n$ - $m$ via the Kirchhoff's current law:

$$\boldsymbol{I}_m = \begin{bmatrix} I_m^a \\ I_m^b \\ I_m^c \end{bmatrix} = \begin{bmatrix} I_{m,in}^a \\ I_{m,in}^b \\ I_{m,in}^c \end{bmatrix} + \sum_{l \in \mathcal{N}(m)} \begin{bmatrix} I_l^a \\ I_l^b \\ I_l^c \end{bmatrix} \quad (4)$$

where $\mathcal{N}(m)$ denotes the set of the downstream branches connected to bus $m$, and $I_l^\varphi$ denotes the sending-end current at phase $\varphi$ of branch $l$.

The sending-end current $\boldsymbol{I}_n$ at branch $n$ - $m$ is calculated based on (4) by

$$\boldsymbol{I}_n = \boldsymbol{c}_{nm} \begin{bmatrix} U_m^a \\ U_m^b \\ U_m^c \end{bmatrix} + \boldsymbol{d}_{nm} \begin{bmatrix} I_m^a \\ I_m^b \\ I_m^c \end{bmatrix} = \boldsymbol{c}_{nm} \boldsymbol{U}_m + \boldsymbol{d}_{nm} \boldsymbol{I}_m \quad (5)$$

where $\boldsymbol{c}_{nm}$ and $\boldsymbol{d}_{nm}$ are determined based on the line parameters at this branch, $\boldsymbol{c}_{nm} = \boldsymbol{Y}_{abc} + \frac{1}{4} \boldsymbol{Y}_{abc} \boldsymbol{Z}_{abc} \boldsymbol{Y}_{abc}$ and $\boldsymbol{d}_{nm} = \boldsymbol{I} + \frac{1}{2} \boldsymbol{Z}_{abc} \boldsymbol{Y}_{abc}$; $\boldsymbol{I}$ denotes an identity matrix.

3) *Forward Sweep:* Starting from the root bus and moving towards the end bus of the feeder, the nodal voltage at bus $m$ is calculated from the voltage at bus $n$ and the sending-end current by

$$\boldsymbol{U}_m = \boldsymbol{a}_{nm} \begin{bmatrix} U_n^a \\ U_n^b \\ U_n^c \end{bmatrix} - \boldsymbol{b}_{nm} \begin{bmatrix} I_n^a \\ I_n^b \\ I_n^c \end{bmatrix} = \boldsymbol{a}_{nm} \boldsymbol{U}_n - \boldsymbol{b}_{nm} \boldsymbol{I}_n \quad (6)$$

where $\boldsymbol{a}_{nm} = \boldsymbol{I} + \frac{1}{2} \boldsymbol{Z}_{abc} \boldsymbol{Y}_{abc}$, and $\boldsymbol{b}_{nm} = \boldsymbol{Z}_{abc}$.

The iterative procedure continues until the voltage differences at all nodes in two successive iterations satisfy

$$\left\| \boldsymbol{U}_k^{(t+1)} - \boldsymbol{U}_k^{(t)} \right\|_\infty < \varepsilon \qquad k = \{1, 2, ..., N\} \quad (7)$$

where $\boldsymbol{U}_k^{(t)}$ and $\boldsymbol{U}_k^{(t+1)}$ denote the three-phase voltages at bus $k$ at iterations $t$ and $t+1$, and $\varepsilon$ denotes the iteration tolerance.

The flowchart of power flow calculation is shown in Fig.2. Furthermore, we calculate the total active power loss in the whole system based on the nodal voltages at all buses by:

$$P_{loss} = \sum_{m:n \to m} \text{real}(\boldsymbol{U}_{nm}^T \boldsymbol{I}_{nm}) \quad (8)$$

where $\boldsymbol{U}_{nm}$ denotes the three-phase voltage drop at branch $n$-$m$, and $\boldsymbol{U}_{nm} = \boldsymbol{U}_n - \boldsymbol{U}_m$; $\boldsymbol{I}_{nm}$ is the three-phase current through the line impedance, and $\boldsymbol{I}_{nm} = \boldsymbol{Z}_{abc}^{-1} \boldsymbol{U}_{nm}$; the function $\text{real}(\cdot)$ takes the real part of the complex number, and $(\cdot)^T$ denotes the transpose of a vector.

III. PROPOSED VVO ALGORITHM

This section integrates the models of switchable capacitor banks, AVRs, and smart inverters into the power flow calculation to evaluate the impacts of the status changes of these regulating devices on distribution system operation. Leveraging these changes as control actions, we propose a multi-agent DQN-based VVO method to realize effective voltage regulation and power loss reduction.

*A. Voltage Regulation Devices*

1) *Capacitor Bank*

We adopt the three-phase model of capacitor banks. Specifically, we define the reactive power provided by the capacitor installed on phase $\varphi$ as a function of the control variable, $a_c^\varphi \in \{0,1\}$, which indicates the status (on/off) of this capacitor. The capacitor provides reactive power when it is connected, i.e., $a_c^\varphi = 1$, and the reactive power at bus $k$ is calculated by the following nonlinear function of $U_k^\varphi$.

$$Q_c^\varphi = a_c^\varphi (U_k^\varphi)^2 B_k^\varphi \quad (9)$$

where $U_k^\varphi$ denotes the $\varphi$-phase voltage of the capacitor installed at bus $k$, and $B_k^\varphi$ denotes the susceptance of the capacitor on phase $\varphi$.

2) *Voltage Regulator*

A 32-step voltage regulator with a regulating range of $\pm 10\%$ is used in distribution systems, and the series and shunt impedance of the voltage regulators are neglected since their values can be regarded extremely small [16]. Define $a_r^\varphi$ as the step for the voltage regulator on phase $\varphi$, and $a_r^\varphi$ takes values between 0.9 and 1.1 at a step of 0.00625 p.u. The control variable for the regulator is defined by

$$a_r^\varphi = \sum_{j=1}^{33} b_j \alpha_j \quad (10)$$

where $\alpha_j$ denotes the binary variable for the $j$th regulator step position, and $\sum_{j=1}^{33} \alpha_j = 1$; $b_j \in \{0.9, 0.90625, ..., 1.1\}$ and $\varphi \in \{a, b, c\}$.

For branch $n$ - $m$ that with the regulator installation, an additional bus $n'$ is introduced between buses $n$ and $m$. The impacts of the regulators installed at this branch on the voltage $\boldsymbol{U}_n$ and current $\boldsymbol{I}_n$ are quantified by [22]

$$\boldsymbol{U}_n{'} = \boldsymbol{A}_r \boldsymbol{U}_n \quad (11)$$
$$\boldsymbol{I}_n{'} = \boldsymbol{D}_r \boldsymbol{I}_n \quad (12)$$

where $\boldsymbol{U}_n{'}$ denotes the three-phase voltage at bus $n'$, and $\boldsymbol{I}_n{'}$ denotes the three-phase current that flows out from this regulator; $\boldsymbol{A}_r = diag\{a_r^a, a_r^b, a_r^c\}$, and $\boldsymbol{D}_r = \boldsymbol{A}_r^{-1}$.

In the power flow calculation, we replace $U_n$ and $I_n$ with $U_n'$ and $I_n'$ at the locations of regulators to run the forward and backward sweeps in (5) and (6).

*3) DG With Smart Inverters*

We develop a per-phase model of smart inverters installed at DGs and assume these smart inverters adopt the reactive power control (RPC) strategy [16]. Assume a DG unit with the smart inverter installed at bus $k$ and the active power of DG outputs is known, and its nominal per-phase capacity is $S_{dg,k}^\varphi$. The $\varphi$-phase reactive power provided or absorbed by this DG unit at bus $k$ can be expressed as the following box constraint:

$$\underline{Q}_{dg,k}^\varphi \leq Q_{dg,k}^\varphi \leq \overline{Q}_{dg,k}^\varphi \tag{13}$$

where $\overline{Q}_{dg,k}^\varphi$ denotes the maximum reactive power of this DG unit installed at bus $k$, and $\underline{Q}_{dg,k}^\varphi = -\overline{Q}_{dg,k}^\varphi$; $\overline{Q}_{dg,k}^\varphi = \sqrt{(S_{dg,k}^\varphi)^2 - (P_{dg,k}^\varphi)^2}$, and $P_{dg,k}^\varphi$ denotes the $\varphi$-phase active power; here we define the control variable as $a_{dg}^\varphi \in [-1,1]$, and $Q_{dg,k}^\varphi = a_{dg}^\varphi \overline{Q}_{dg,k}^\varphi$. The dispatchable range of $Q_{dg,k}^\varphi$ is relatively narrow since a high power factor (*e.g.*, 0.95) is preferable during DG operation.

To introduce control strategies of smart inverters into the DQN-based VVO framework, we discretize the action space of $a_{dg}^\varphi$ to handle the performance of these actions in RL. This processing method is widely accepted and used for flexible Q-learning or DQN applications, such as [23] and [24]. Here, we suppose that DG opeartors have a certain number of strategies for each control interval (13) in practice [25], and $a_{dg}^\varphi$ takes values between $-1$ and 1 at a step of 0.1.

Fully considering all setting changes of the capacitor banks, smart inverters, and AVRs, the power flow calculation process in (3)-(7) is updated by integrating (9)-(13), along with the voltage-dependent loads modeled by (1) and (2). Moreover, under time-varying operating conditions, the effective power flow calculation acts as the environment for DQN agents, and the details of the proposed VVO algorithm are shown in the next section.

*B. Multi-agent DQN-based Method*

In a DQN-based RL process, a NN is defined as an agent, and the part where the agent takes control actions is the environment. Massive episodes of training are applied to the agent, and in the environment, the load consumption and DG production in a distribution system vary in each episode. The DQN agent is required to take control actions with respect to the given operating condition to achieve VVO. The dimension of the action space explosively increases with the number of controllable devices installed in the three-phase distribution system. Also, a single-agent DQN is challenging to efficiently provide actions due to the extremely high dimension of the joint action space [21]. To improve computational efficiency and ensure scalability for VVO, we propose a multi-agent DQN-based algorithm. The interaction between multiple agents and the environment is depicted via three elements: state $s$, action

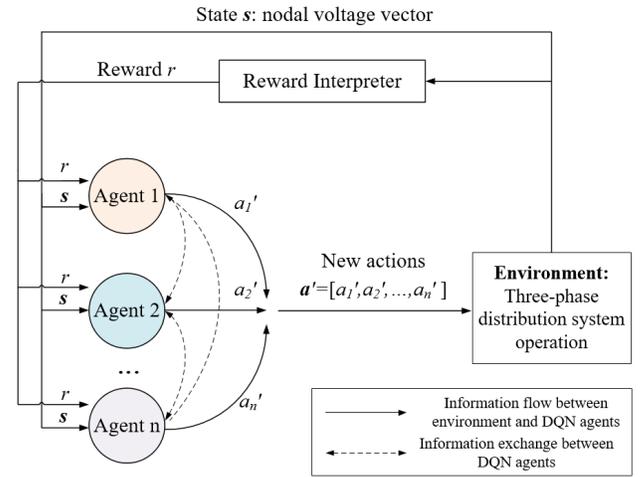

Fig.3. Multi-agent DQN training loop

$a$, and reward $r^t$ at episode $t$. We describe these elements and the offline training and online test process for VVO below.

*1) State and Action*

The multi-agent DQN distributes the global control actions to each agent that performs coordinated RL by exploring the shared environment [21]. Here, let $j$ denote the serial number of an agent, and $j \in \{1,2,\dots,n\}$, where $n$ is the number of the agents. Considering various regulating devices, the actions of all the devices are assigned to multiple agents, shown in Fig. 3. The action vector for the VVO problem decided by all agents is expressed as

$$\boldsymbol{a} = [a_1, a_2, \dots, a_n]^T \tag{14}$$

where $a_j$ denotes the control action of agent $j$ that originates from the statuses of three-phase smart inverters, regulators, or capacitors, *i.e.*, $a_r^\varphi$, $a_c^\varphi$, and $a_{dg}^\varphi$; $a_j \in \mathcal{A}_j$, and $\mathcal{A}_j$ represents the searching space of the corresponding action for the devices in agent $j$.

For the action vector $\boldsymbol{a}$ provided by the agents, the environment provides three-phase voltages at all buses in distribution systems, which act as the states of the DQN. Moreover, these states are expressed as $\boldsymbol{s} = [\boldsymbol{U}_1, \boldsymbol{U}_2, \cdots, \boldsymbol{U}_N]$, where $\boldsymbol{U}_k = [U_k^a, U_k^b, U_k^c]$ and $k = \{1, 2, \dots, N\}$.

*2) Reward*

The objective of VVO used here is to keep the nodal voltages in a normal range (*e.g.*, 0.95 to 1.05 p.u.) and concurrently minimize total active power loss [11], [12]. Hence, the optimization model for VVO can be formulated as

$$\boldsymbol{a} = \arg\min\ P_{loss}^t \tag{15.a}$$
$$\text{s.t.}\ \ 0.95 \leq |U_k^\varphi| \leq 1.05 \tag{15.b}$$

where $|U_k^\varphi|$ denotes the voltage magnitude on phase $\varphi$ at bus $k$, and $\varphi = \{a,b,c\}$; $P_{loss}^t$ is calculated by (8) at episode $t$.

To solve the optimization model (15), we construct the reward interpreter of DQN by putting the constraints into the objective function. The reward interpreter has the following characteristics:

▪ When the voltage constraint (15.b) is not violated, we calculate the reward value at episode $t$ by

$$r^t = P_{loss}^{t-1} - P_{loss}^t \tag{16}$$

where $P_{loss}^t$ denotes the active power loss based on the current action variables, and $P_{loss}^{t-1}$ denotes the one that takes the previous actions at episode $t-1$, both of which are calculated by (8) according to the corresponding states; $r^t > 0$ implies that the proposed DQN further reduces the power loss after conducting the new action given by the agents at episode $t$.

- If the constraint (15.b) is violated, a significant penalty $M$ is imposed to the reward of the DQN, which moves the voltages into the normal range [11]. To accurately quantize the voltage deviation degree in the whole distribution system, the reward function in the case of voltage violation is calculated by

$$\begin{aligned} r^t = -M \sum_{\varphi} \sum_k & [\max(|U_k^{\varphi}| - 1.05, 0) \\ & + \max(0.95 - |U_k^{\varphi}|, 0)] \end{aligned} \tag{17}$$

where the more significant the degree of voltage violation is, the more negative reward the DQN agents obtain.

*3) Offline Training and Online Test*

In the proposed multi-agent DQN, efficient communication among these agents is conducted to select the optimal actions via their shared observation of the current state $s$, shown as Fig. 3. Moreover, the information exchanged among the agents is the current actions that these agents jointly take, $a$. In each training episode, based on the current state $s$, the agents provide control actions to the environment. Collect each control action from all agents to form the new action vector for the distribution system, $a' = [a_1', a_2', \ldots, a_n']^T$. The environment, *i.e.*, the power flow calculation procedure, then implements the joint action $a'$ and get a new reward and a new state $s'$, until the training process terminates.

According to our customized reward function (16) and (17), the new state $s'$ and the corresponding system power loss are interpreted into the immediate reward $r$ after taking action $a$ at state $s$. For $t = 1, 2, \ldots, N_{ep}$, agent $j$ updates the action-reward $Q$ function at episode $t$ via the following Bellman equation:

$$Q_j^{t+1}(s,a) = Q_j^t(s,a) + \alpha(r^t + \gamma \max Q_j^t(s',a') - Q_j^t(s,a)) \tag{18}$$

where $\gamma \in [0,1]$ is a discount rate, and $\alpha$ denotes the learning rate of the DQN.

An *experience replay* technique is used to store the latest $N_b$ sets of the agents' experience in episode $t$, *i.e.*, the transition tuple $(s, a, r^t, s')$, to a replay buffer $\mathcal{M}$. We sample a mini-batch memory $\mathcal{D}$ from the replay buffer to improve the generality of the agents towards diverse states. The agent $j$ is trained by $\mathcal{D}$ together with the current transition tuple. Moreover, the stochastic gradient descent on NN parameters $\theta_j$ for the agent is conducted using the following loss function $\mathcal{L}_j(\theta_j)$, which enforces the Bellman equation (18):

$$\mathcal{L}_j(\theta_j) = \mathbb{E}\left[\left(r^t + \gamma \cdot \max Q_j^t(s',a') - Q_j^t(s,a)\right)^2\right] \tag{19}$$

where we define the target $Q$ function as $y = r^t + \gamma \cdot \max Q_j^t(s',a')$, and $\mathbb{E}(\cdot)$ denotes the expectation function.

The $Q$ function iteratively updates following (18), shown as $Q_j^{t+1}(s,a) = \mathbb{E}(r^t + \gamma \max Q_j^t(s',a'))$. Such iterations

---

**Multi-agent DQN Training Process**

| | |
|---|---|
| 1 | **Input**: Distribution system model and the action space $\mathcal{A}_j$ for agent $j$, $j \in \{1,2,\ldots,n\}$. |
| 2 | **Initialization**: the learning rate $\alpha$, the discount rate $\gamma$, the decay factor $\eta$, and the size of replay buffer $N_b$. |
| | **for** $t = 1$ to $N_{ep}$, **do** |
| 3 | Initialize state $s$, and obtain action $a$ by the $\varepsilon$-greedy policy (20). |
| | **for** $j = 1$ to $n$, **do** |
| 4 | Get reward $r^t$ by (16) and (17), and new state $s'$ by power flow calculation, and store them as a transition $(s, a, r^t, s')$ into a replay buffer. |
| 5 | Get the current $Q$ vector at state $s$ by agent $j$. |
| 6 | Sample from the replay buffer to obtain tuple $(s(i), a(i), r^t(i), s'(i))$, and $i = 1, \ldots, N_d$. |
| 7 | Set $y = r^t + \gamma \max Q_j^t(s',a')$ |
| 8 | Train and update agent $j$ by performing gradient descent on (19). |
| | **end** |
| | **end** |
| 9 | **Output**: All agents with parameters $\theta_j^*$. |

converge to the optimal action-value function, $Q_j^t \to Q_j^*$ as $t \to \infty$ [26].

During the training process, we apply the $\varepsilon$-greedy policy [20] to select the actions efficiently, as it encourages each agent to fully explore the corresponding action space. Specifically, as the training continues, the action selection relies more on the action policy from $Q_j^t(s',a')$, shown as:

$$a_j = \pi_j(s) = \begin{cases} \text{random action from } \mathcal{A}_j, & \text{if } \xi < \varepsilon_t \\ \arg\max_{a_j' \in \mathcal{A}_j} Q_j^t(s',a'), & \text{otherwise} \end{cases} \tag{20}$$

where $\pi_j$ denotes the action selection policy for agent $j$, and $0 < \xi < 1$ is a random number; the searching criteria $\varepsilon_t$ is updated based on the last episode by a decay factor $\eta$, *i.e.*, $\varepsilon_t = \varepsilon_{t-1}\eta$.

The pseudo-code summarizes the offline training process of the proposed MADRL algorithm. When the training process terminates, the agent $j$ with parameter $\theta_j^*$ is applied to the test cases, where new operating conditions in the distribution system are fed. For each test case, these well-trained agents provide the action policy by

$$a_j = Q_j^*(s, a; \theta_j^*) \qquad j \in \{1,2,\ldots,n\} \tag{21}$$

These actions from all agents are combined by (14) and given to the environment as the solution of the model (15) for online system VVO control.

## IV. CASE STUDY

We test the proposed algorithm on the three-phase unbalanced IEEE 13-bus and 123-bus distribution systems [27]. We modify the 13-bus system by adding two single-phase PV units at buses 675 and 684, and a three-phase PV unit at bus 680, illustrated in Fig. 4. Six DG units are added at buses 13, 18, 60, 151, 250, and 300 in the 123-bus systems, and the installation capacity of these DG units is set as 300 or 500 kVA [28]. Table I lists the specifications of control devices including smart inverters installed at the DG units in both test systems. We see from Table I that taking the action variables for all these devices into account results in a high-dimensional joint decision space. For offline training and online test, we randomly generate 9,000 and 13,000 operating conditions (episodes) with 80% to 120% of random fluctuations of base loads in these two systems, respectively [20]. Moreover, in the test phase, new operating conditions are used as test cases, and Table II gives the number of training episodes and test cases, $N_{ep}$ and $N_{test}$.

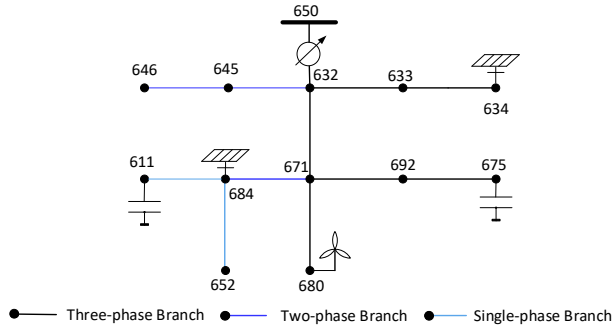

Fig. 4. Three-phase unbalanced 13-bus distribution system

TABLE I
INSTALLATION SETTINGS OF CONTROL DEVICES

| Device Type | 13-bus System | | 123-bus System | |
|---|---|---|---|---|
| | No. bus/branch | Phase | No. bus/branch | Phase |
| Regulator | 650-632 | A, B, C | 150-149 | A, B, C |
| | | | 9-14 | A |
| | | | 25-26 | A, C |
| | | | 60-67 | A, B, C |
| Capacitor | 611 | C | 83 | A, B, C |
| | 675 | A, B, C | 88 | A |
| | | | 90 | B |
| | | | 92 | C |
| Smart Inverter | 684 | C | 13 | C |
| | | | 18 | A, B, C |
| | | | 60 | A, B, C |

TABLE II
PARAMETER SETTING OF MULTI-AGENT DQN

| Hyperpara. | 13-bus System | 123-bus System |
|---|---|---|
| $\eta$ | 0.999 | 0.999 |
| $\gamma$ | 0.95 | 0.95 |
| $N_b$ | 2000 | 5000 |
| $N_{ep}$ | 8000 | 9000 |
| $N_{test}$ | 1000 | 4000 |

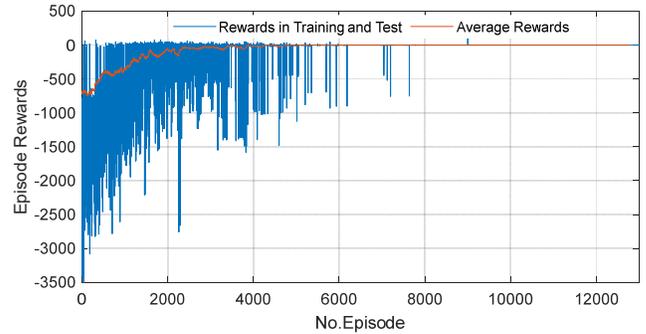

Fig.5. Multi-agent DQN training process in the IEEE 123-bus system

*DQN Specification:* The NNs used here have three fully connected layers, and the learning rate is chosen as 0.0001. The NN agents in the proposed method use rectified linear unit (ReLU) activation functions in the hidden layers and sigmoid functions in the output layer. Table II also summarizes the hyperparameters of the adopted DQN in the two test systems.

*A. Learning Performance*

We investigate the learning performance of the proposed method. We test the proposed algorithm on the 123-bus system, and Fig. 5 shows the reward values in the training process and average rewards in successive 200 episodes. We observe that the DQN agents' control policies result in negative rewards ($r^t < 0$) due to limited positive learning experiences and not yet optimized action policies in an earlier learning phase. These negative rewards illustrate that at the beginning, the action policies are incapable of maintaining the system voltages within 0.95~1.05 p.u. at all times, according to (16) and (17). However, during the training process, the agents gradually evolve and obtain positive rewards ($r^t > 0$) more and more frequently. Moreover, $r^t > 0$ implies that there is no voltage violation and the power loss is further reduced by taking action $a'$, compared with the performance before $a'$ is taken. Also, the average reward in the training process continuously increases, which shows the DQN's ability in realizing the multi-objective VVO.

Furthermore, for online test, 4,000 new cases are fed to these trained agents. These agents demonstrate the effective control performance for VVO, which is characterized by the positive rewards in these cases, as shown in episodes from 9,000 to 13,000 of Fig. 5. We conclude that the proposed DQN enables the power grid to self-learn with the "cognitive" function of VVO control by mimicking the human mind. Eventually, these trained agents can implement effective control policies when confronted with new operating conditions.

*B. VVO Performance*

This section demonstrates the VVO performance of the proposed model-free MADRL method when facing random operating conditions, in terms of voltage regulation and power loss reduction, both of which are implemented online by this method simultaneously.

*1) Voltage Regulation*

To evaluate the voltage control performance of the proposed VVO method in the test cases, we define the success rate in voltage regulation, $S_v$, as follows.

$$S_v = N_v/M_v \quad (22)$$

where $M_v$ denotes the number of those test cases that exist voltage violation before adopting the proposed VVO algorithm, and $N_v$ denotes the number of those cases that avoid the voltage violation issue after adopting this method. Moreover, a higher $S_v$ illustrates that the proposed algorithm has a better control performance in voltage regulation.

Table III summarizes the statistical results of the proposed algorithm in test cases. Specifically, before employing the proposed method in the 13-bus system, voltage violation exists in the 1,000 test cases, and our algorithm achieves a control success rate of 99.80% in voltage regulation. For a larger-scale three-phase 123-bus system, the VVO task becomes more challenging due to the so-called "curse of dimensionality." However, the proposed MADRL method obtains a success rate that reaches up to 99.975%.

TABLE III
PERFORMANCE OF VOLTAGE REGULATION

| System Scale | $N_p$ | $M_p$ | $S_p$ |
|---|---|---|---|
| 13-bus System | 998 | 1,000 | 99.80% |
| 123-bus System | 3,999 | 4,000 | 99.975% |

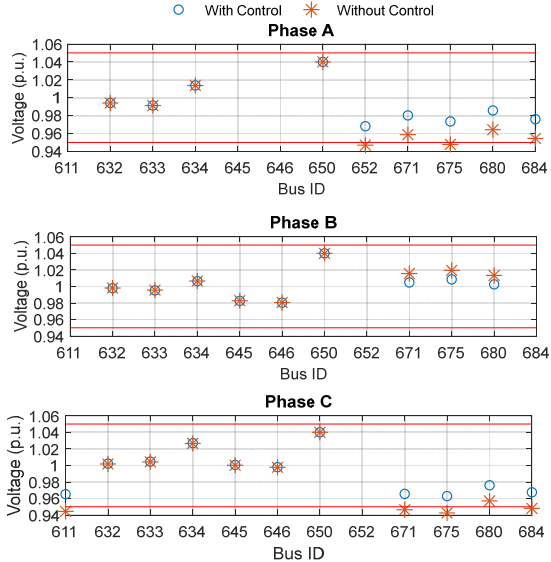

Fig.6. Three-phase voltage magnitude comparison before and after control in the 13-bus system.

TABLE IV
VVO SETTING RESULTS IN THE IEEE 123-BUS SYSTEM

| VVO Control Policy | Phase A | Phase B | Phase C |
|---|---|---|---|
| Reg. 1 Tap | 21 | 21 | 21 |
| Reg. 2 Tap | 10 | - | - |
| Reg. 3 Tap | 13 | - | 13 |
| Reg. 4 Tap | 22 | 22 | 22 |
| Cap. 1 Status | ON | ON | ON |
| Cap. 2 Status | ON | - | - |
| Cap. 3 Status | - | ON | - |
| Cap. 4 Status | - | - | ON |
| Smart Inv. 1 $Q_{dg,13}^\varphi$ | - | - | 35.31 kVAR |
| Smart Inv. 2 $Q_{dg,18}^\varphi$ | 49.67 kVAR | 90.62 kVAR | 52.58 kVAR |
| Smart Inv. 3 $Q_{dg,60}^\varphi$ | 6.21 kVAR | 11.34 kVAR | 6.58 kVAR |

TABLE V
VOLTAGE COMPARISOM IN THE IEEE 123-BUS SYSTEM

| Voltage Magnitude Profiles | | Phase A | Phase B | Phase C |
|---|---|---|---|---|
| Without VVO Control | Min. Voltage [p.u.] | 0.7243 | 0.7025 | 0.7361 |
| | Max. Voltage [p.u.] | 1.0300 | 1.0300 | 1.0300 |
| | Avg. Voltage [p.u.] | 0.7973 | 0.8063 | 0.8218 |
| With VVO Control | Min. Voltage [p.u.] | 0.9799 | 0.9882 | 0.9774 |
| | Max. Voltage [p.u.] | 1.0403 | 1.0427 | 1.0422 |
| | Avg. Voltage [p.u.] | 0.9965 | 0.9999 | 1.0101 |

We randomly select a test case in the 13-bus system to compare the voltage magnitudes without and with control, and Fig. 6 depicts these three-phase voltage magnitudes at each bus. It can be seen that without control, the A-phase and C-phase voltage magnitudes at buses 611, 652, 671, 675, and 684 violate the normal voltage operation limits; after adopting the control actions provided by the agents, the voltages at all these buses fall within the normal operating range. For the 123-bus system, the VVO results of the proposed method in a test trial, *i.e.*, the taps/statuses of the regulating devices, are shown in Table IV. Moreover, we compare the voltage profiles with and without VVO control in the test case, in terms of the minimum, maximum, and average of voltage magnitudes, in Table V. It is shown that the proposed method avoids voltage violation by jointly dispatching these devices, and these DG units with smart inverters are directed to provide reactive powers for voltage lift.

*2) Power Loss Reduction*

Here we also show the performance of power loss reduction in those test cases that are discussed above for voltage regulation. Fig. 7 demonstrates the power loss comparison without and with control in 50 test cases that are randomly selected in the 13-bus system. We conclude that that the proposed method enables effective power loss reduction.

To further quantify the loss reduction performance in these two test systems, we calculate the reduction of power loss at each test case by

$$\Delta P^i = P_{loss,0}^i - P_{loss}^i \qquad (23)$$

where $\Delta P^i$ denotes the difference in $i$th test case between the active power losses with and without the control strategy suggested by the proposed method, *i.e.*, $P_{loss,0}^i$ and $P_{loss}^i$; $i = 1, 2, \ldots, N_{test}$, and these active power losses are calculated by (8). Here, we assume that before taking the control actions, no capacitor bank is connected in these systems, and the steps of AVRs and the power factors of smart inverters are set as 1 as default [29].

Table VI summarizes the average and maximum of $\Delta P^i$ in two test systems and implies that the proposed method enables power loss reduction in all test cases. Specifically, in the 13-bus system, the average power loss reduction obtained by the proposed algorithm is 34.12 kW and averagely accounts for 14.78% of the power loss without control. In the 123-bus system, these statistic data are 109.09 kW for the mean power loss reduction and 36.09% for the loss reduction percentage. We conclude that the proposed DQN-based method effectively realizes dual goals on power loss reduction and voltage regulation simultaneously.

*C. Computation Time*

We carry out numerical experiments to investigate the computational efficiency of the proposed method. All the tests are performed using MATLAB on the machine equipped with a 2.5 GHz Intel Core i5 CPU and 8 GB of RAM.

In the online test phase, the average executive time of all test cases in the 13-bus and 123-bus systems is 21.7 and 39.2 milliseconds, respectively, which is promising to meet with the requirement of real-time implementation in power systems. The proposed algorithm still shows the high computational efficiency in the unbalanced 123-bus system. Moreover, the proposed MADRL method is competitive when dealing with a high-dimensional action space that exists in three-phase distribution systems.

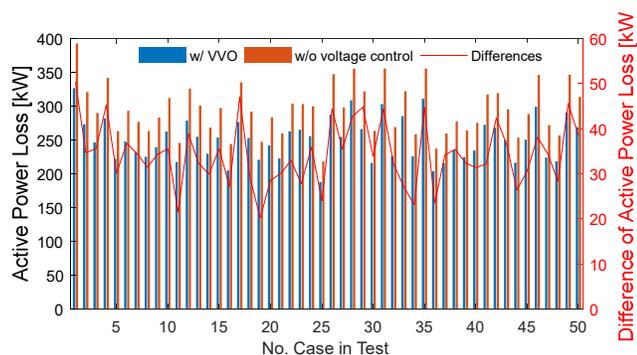

Fig.7. Power loss comparison before and after control in the 13-bus system.

TABLE VI
PERFORMANCE OF POWER LOSS REDUCTION

| System Scale | Average $\Delta P^i$ [kW] | Maximum $\Delta P^i$ [kW] | The prop. of $\Delta P^i > 0$ |
|---|---|---|---|
| 13-bus System | 34.12 | 62.22 | 100% |
| 123-bus System | 109.09 | 123.55 | 100% |

## V. CONCLUSION

This paper proposes a novel and real-time DQN-based VVO algorithm in unbalanced distribution systems. Integrating the voltage-dependent loads, DG penetration, and three types of voltage regulating devices into distribution system operation, we establish the efficient power flow calculation as the environment of the DQN. Via the interaction between the environment and multiple agents, the proposed VVO method adaptively chooses control actions to enable voltage regulation and power loss reduction. This algorithm realizes a promising VVO performance in two unbalanced distribution systems. Future work focuses on the implementation of MADRL in load frequency control via adopting continuous variables as actions in three-phase unbalanced distribution systems.

Submitted in 2019

ulating -Voltage -report.pdf